\documentclass[12pt,preprint]{aastex}
\def\p/{\mbox{$^1$}}
\def\pp/{\mbox{$^2$}}
\def\ppp/{\mbox{$^3$}}
\def\pppp/{\mbox{$^4$}}
\def\m/{\mbox{$^{-1}$}}
\def\mm/{\mbox{$^{-2}$}}
\def\mmm/{\mbox{$^{-3}$}}
\def\mmmm/{\mbox{$^{-4}$}}
\def\Ms/{\mbox{M$_\odot$}}

\slugcomment{Submitted to Astrophysical Journal Supplements}
\begin{document}

\title{A Galactic O-Star Catalog}
\shorttitle{A Galactic O-Star Catalog}

\author{Jes\'us Ma\'{\i}z-Apell\'aniz\altaffilmark{1}}
\affil{Space Telescope Science Institute\altaffilmark{2}, 3700 San Martin 
Drive, Baltimore, MD 21218, U.S.A.}

\author{Nolan R. Walborn}
\affil{Space Telescope Science Institute\altaffilmark{2}, 3700 San Martin 
Drive, Baltimore, MD 21218, U.S.A.}

\author{H\'ector \'A. Galu\'e}
\affil{Instituto Tecnol\'ogico y de Estudios Superiores de Monterrey, Av.
Eugenio Garza Sada 2501, 64849 Monterrey, Nuevo Le\'on, M\'exico}

\author{Lisa H. Wei}
\affil{Department of Astronomy, Cornell University, Ithaca, NY 14853, U.S.A.}

% ---------------affiliations of the science group -----------------------

\altaffiltext{1}{Affiliated with the Space Telescope Division of the European 
Space Agency, ESTEC, Noordwijk, Netherlands.}
\altaffiltext{2}{The Space Telescope Science Institute is operated by the
Association of Universities for Research in Astronomy, Inc. under NASA
contract No. NAS5-26555.}

\begin{abstract}
	We have produced a catalog of 378 Galactic O stars with accurate
spectral classifications which is complete for $V<8$ but includes many
fainter stars. The catalog provides cross-identifications with other sources;
coordinates (obtained in most cases from Tycho-2 data); astrometric distances
for 24 of the nearest stars; optical (Tycho-2, Johnson, and Str\"omgren) and
NIR photometry; group membership, runaway character, and multiplicity 
information; and a web-based version with links to online services.
\end{abstract}


\section{Introduction}

	The advent of internet search engines and databases is producing a
revolution in astronomy. Nowadays, it is possible to use a number of web-based
tools to extract spectroscopic, photometric, kinematic, or astrometric data from
almost any source published in the last century. Also, new large-scale surveys
are making huge amounts of data available to any astronomer sitting in front of
a computer screen connected to the internet. This has made possible a new kind
of astronomer, the electronic one, who can work with previously unanalyzed data
and combine it with information from search engines to produce interesting
science without ever visiting an observatory.

	However, behind the possibilities offered by this avalanche of
information lurks a danger. When one compares data from widely diverse sources, 
the possibility exists that apples and oranges are being mixed: good quality
results with poorer ones, data with different systematic errors, or measurements
obtained with different calibrations (filter sets, observatories, data 
reduction packages\ldots). One of the fields where this danger is more
apparent is spectral classification, a subject where lower quality
data or the technique of the observer can easily introduce systematic errors. 
It is in those cases where a critical compilation of a catalog using criteria
as uniform as possible becomes important, since different sources can provide
quite different classifications. As an example of the problems in this field, 
if one searches in Simbad for the
spectral classification of $\delta$ Ori A, one of the brightest O stars in the
sky, 41 references are returned with spectral classes between O8 and O9.5 and
luminosity classes between III and Iab. $\theta^1$ Ori A, one of the stars in
the Trapezium, provides an even more dramatic example: the two results obtained 
in Simbad are O7 and O8 V n; in reality it is an early-B star (see next 
section).

	A number of lists and catalogs of spectral types for O stars are
available in the literature. The works of \citet{Morgetal55}; \citet{Hilt56};
and \citet{Lesh68}
provided original accurate spectral classifications for ``northern'' 
($\delta \gtrsim -20\degr$) stars which, taken together, constitute a
quite uniform catalog for that part of the sky. The southern equivalent can be
obtained by merging the spectral classifications obtained by \citet{Hiltetal69}
and \cite{Garretal77}. Those catalogs, though, are far from complete
and most of their spectral classifications were done before the introduction
of spectral types O3 \citep{Walb71b} and O2 \citep{Walbetal02b}. A different
approach was taken by \citet{Goy80} and by \citet{Garmetal82}, who built 
catalogs of 970 and 765 stars, respectively, by compiling spectral 
classifications from different sources (39 in the first case, 8 
in the second one). These catalogs are impressive for its number of stars but, 
unfortunately, its spectral classifications are far from uniform and complete.
This is reflected in \citet{Goy80} having multiple spectral classifications for
many of its stars, and in \citet{Garmetal82} with many stars not having
luminosity classes and with a non-negligible number of errors (see
next section for a few of them). Another alternative is to use objective
prisms to perform large-scale surveys, of which the most comprehensive one is
the ongoing Michigan Spectral Catalog \citep{HoukSwif99}. The resolution of
those surveys tends to be lower and the spectral classifications can have
systematic errors with respect to studies performed with slits (in the Michigan
catalog a number of stars classified as O are actually of early-B type), making
the former less precise. Finally, a recent catalog \citep{Reed03} takes the
broad approach of compiling ``OB'' (meaning O to approximately B2) stars by
adding stars on the basis of optical photometry alone. Given good-quality
photometry, this method is capable of increasing the sample considerably 
(\citealt{Reed03}, indeed, contains over $16\, 000$ stars) but, given the 
color degeneracy existent between O and early B stars, it is not of much use to
distinguish between them and, given the relative numbers of the two samples,
one would expect that most of the stars in that catalog are actually of
early-B type.

	Our approach in this article will be to build a catalog of O stars with
the following conditions. (1) Stars should be selected on the basis of their
optical spectroscopy alone. (2) Sources for spectral classification should be
accurate and as uniform as possible. (3) Valuable additional information
(positions, photometry, distances\ldots) should be added only if they come from
uniform catalogs or if their accuracy can be checked. (4) The catalog should be
made accessible through the web and be actively maintained in order for updates 
to reach the users as fast as possible. With those criteria in mind, we intend 
this article to be just the first version of the Galactic O-star (GOS) catalog.
In the future, if resources are found, we intend to add data from other 
missions, such as UV photometry and spectroscopy from GALEX and IUE; 
expand the catalog to include more
stars, increasing the completeness limit in magnitude; and collect more data,
e.g. producing a spectroscopic library. Updates (and possible errata) will be
posted at the web site {\tt http://www.stsci.edu/\~{}jmaiz/GOSmain.html}.

	This paper is organized as follows. In section 2 we describe how the
catalog was constructed. In section 3 we lay out the criteria used for the
spectral classification. In sections 4, 5, and 6 we describe how the astrometry, the
photometry, and other information in the GOS catalog was compiled, respectively.

\section{Catalog construction, nomenclature and cross-identification}

	The GOS catalog was built initially by collecting all 
Galactic O-spectral-type classifications produced by one of the authors (N.R.W.)
over the years in the following primary references:
\citet{Walb72,Walb73,Walb74,Walb76,Walb80,Walb81,Walb82,Drisetal95,WalbFitz00,WalbHowa00};
and \citet{Walbetal02b}, 
yielding a total of 350 stars. We also included other known O stars with $V<8$
using the following procedure. We first selected the five major catalogs 
quoted in the Introduction, which include a large number of O stars and 
which were compiled using uniform spectral classification criteria: 
\citet{Morgetal55,Hilt56,Lesh68,Hiltetal69}; \and \citet{Garretal77}. 
The search of those
five secondary references yielded 18 additional O stars with $V<8$, which were 
included in the GOS catalog. We then used as tertiary references four 
additional general catalogs \citep{Garmetal82,Gies87,Masoetal98,HoukSwif99} 
plus a number of papers on individual clusters, associations, or stars
\citep{MacCBide76,NiemMorr88,Smitetal90,Willetal90,MassThom91,GarmSten92,MassJohn93,Hilletal93,Browetal94,Massetal95,Burketal97,deMaSchm99,deZeetal99,Schwetal99,Hoogetal01,Pennetal02}
to find additional cases. In order to ensure an adequate degree of uniformity,
we included in the GOS catalog additional O stars with $V<8$ from those 
references only if they were not classified as B stars in one of the primary or 
secondary references. Each case was studied in detail; conflicts are described 
later in this section. 10 O stars were
added from those tertiary references, yielding a total of 378 stars in this
version of the GOS catalog. Therefore, in the absence of errors, we expect our 
catalog to be complete for $V < 8$ unless an O star of that magnitude or 
brighter has escaped detection in all of the above references.

	The stars in the GOS catalog are listed in 
Tables~\ref{GOScatalog_namspcl}~and~\ref{GOSsupplement_namspcl}. The main
catalog (Table~\ref{GOScatalog_namspcl}) lists the majority of the stars, while
the supplement (Table~\ref{GOSsupplement_namspcl}) lists those O stars which 
have unresolved WR companions.
Within each table stars are sorted by Galactic longitude. 
The first column includes the
designation in this catalog (e.g. the first star in 
Table~\ref{GOScatalog_namspcl} is GOS G$001.94-02.62\_01$), which was produced
according to the criteria recommended by the IAU \citep{Lortetal94}.  The first
character is always a G (indicates Galactic coordinates), the next six express 
the Galactic longitude in degrees in the format {\tt LLL.ll}, the eighth 
character is the latitude sign, the next five express the absolute value of the 
Galactic latitude in degrees in the format {\tt BB.bb}, the fourteenth is an 
underscore, 
and the fifteenth and sixteenth are a running number to differentiate stars 
which would otherwise have the same catalog number\footnote{Note that a previous
web-based version of this catalog used a different nomenclature with arcminutes
instead of decimal degrees.}. Each of the two tables is sorted hierarchically
by GOS number, i.e. by longitude, latitude sign, latitude, and running number
(in this order). The second column in 
Tables~\ref{GOScatalog_namspcl}~and~\ref{GOSsupplement_namspcl} lists the
corresponding HD number (when it exists) and the third column the capital
letter and/or numerical identifications  
for the members of a multiple system. Most of the
original spectral classifications identified the stars by their HD numbers, so
in order to facilitate cross-identification we have generated 
Table~\ref{GOShdnumbers}, where all stars in the catalog with HD numbers are
sorted using that identification as sorting criterion.

	As we shall see later, we use the Tycho-2 catalog \citep{Hogetal00a}
as our main source for positions. In order to find the correspondence between
HD and Tycho-2 identifications we utilized as our primary source the work by
\citet{Fabretal02}. Those cases without HD numbers were manually identified as 
Tycho-2 stars using Aladin \citep{Bonnetal00}. We also checked a 
number of other cases manually and we found a few errors or incomplete 
assignments in the HD-Tycho-2 identification table of \citet{Fabretal02}, which
we corrected for in our catalog:

\begin{itemize}
  \item HD 5005 A and C have independent Tycho-2 numbers, but only HD 5005 A
  	is listed in the HD to Tycho-2 identification table.
  \item HD 92206 A and C have independent Tycho-2 numbers but only HD 92206 A
  	is listed. Note that the original spectral classifications in 
	\citet{Walb82} refers to the composite spectrum of AB, and C
	separately.
  \item The Tycho-2 assignment for HD 93128 corresponds in reality to HD 93129
  	A. HD 93128 and HD 93129 B are not Tycho-2 stars.
  \item HD 113904 A and B have independent Tycho-2 numbers but only HD 113904 A
  	is listed in the HD to Tycho-2 identification table.
  \item HD 229232 is listed as Tyc 3152-00569-1 but in reality it is Tyc
        3152-00289-1.
  \item HD 303308 is listed as Tyc 8626-02809-1 but in reality it is Tyc
        8526-02808-1
  \item The assignment for HD 319703 corresponds to the A component of this
  	system.
\end{itemize}

	The resulting Tycho-2 identification is listed in the fourth column of
Tables~\ref{GOScatalog_namspcl}~and~\ref{GOSsupplement_namspcl}. The fifth
column gives the identification from the August 2002 version of the 
Washington Double Star (WDS) catalog \citep{Masoetal01}. The sixth column gives
the most common name for those stars that do not have an HD catalog number or
which are usually known by a name other than their HD designation.

	We end this section by noting that some early-B stars are identified as 
O stars erroneously in the literature. Here we provide a non-comprehensive list 
of the most notorious cases:

\begin{itemize}
  \item {\bf $\theta^1$ Ori:} Only $\theta^1$ Ori C is an O star; the other 
  	three classical members of the Trapezium are early-type B stars
	(see, e.g. \citealt{Schuetal01}). The source of the confusion is 
	that some authors have assigned letters to the the four stars in order 
	of right ascension and others in order of brightness. Note that the
	spectral type assignments in the current printed edition of the
	Bright Star Catalog are incorrect.
  \item {\bf HD  37743:} $\zeta$ Ori B (A is HD 37742) is actually an 
  	early-type B giant but is listed in \citet{Garmetal82} as O9.5~IV.
  \item {\bf HD  61827:} \citet{Garretal77} give B3~Iab (adopted) but 
  	\citet{HoukSwif99} give O8/9~Ib:.
  \item {\bf HD  69106:} \citet{Garretal77} give B0.5~IV (adopted). This is an 
  	example of the archaic HD type Oe5 (between late-type O supergiant and 
	B0) and \citet{Garmetal82} give O5~V.
  \item {\bf HD  75821:} \citet{Morgetal55} give B0~III (adopted) but 
  	\citet{Garmetal82} give O9.5~II.
  \item {\bf HD  91452:} \citet{Garretal77} give B0~Ia (adopted) but in 
  	\citet{Hump73} it is listed as O9.5~Iab-Ib.
  \item {\bf HD  93208:} \citet{Garretal77} give B2~II-III (adopted) but 
  	\citet{HoukSwif99} give O9/B0, while Simbad gives O9.5. Also, there is 
	a possible confusion with HD 93028, which is a real O star.
  \item {\bf HD 116781:} \citet{Garretal77} give B0~IIIne (adopted); 
  	\citet{HoukSwif99} give O9/B1~(I)e.
  \item {\bf HD 122879:} \citet{Walb76} gives B0~Ia (adopted) but it appears as 
        O9.5~Ia in \citet{Garmetal82}.
  \item {\bf HD 151564:} \citet{Garretal77} give B0.5~V (adopted) but 
  	\citet{Garmetal82} give O9.5~IV.
  \item {\bf HD 155775:} \citet{Garretal77} give B0.5~III (adopted); 
  	\citet{Garmetal82} give O9.5.
  \item {\bf HD 161807:} \citet{Garretal77} give B0:~III:nn (adopted) but it 
  	appears as O9.5~V:n in \citet{HoukSwif99}.
  \item {\bf HD 163181:} An example of the archaic HD type Oe5 (between 
  	late-type O supergiant and B0). It is actually BN0.5~Iae var 
	\citep{Walb76} but appears as O5 in \citet{Garmetal82}.
  \item {\bf HD 165319:} \citet{Morgetal55} give B0~Ia (adopted) but it is 
  	classified as O9.5~Iab in \citet{HoukSwif99}.
  \item {\bf HD 191139:} \citet{Walb71a} gives B0.5~II (adopted); 
  	\citet{Garmetal82} give O9.5~II.
  \item {\bf HD 204827:} \citet{Morgetal55} give B0.5~V but \citet{Garmetal82} 
  	give O9.5~V. New data obtained by C. Robert et al. (private 
	communication) yield B0.2~V.
  \item {\bf HD 206183:} \citet{Morgetal55} give B0~V but \citet{Garmetal82} 
  	give O9.5~V. New data obtained by C. Robert et al. (private 
	communication) confirm the B0~V classification.
  \item {\bf HD 207538:} \citet{Morgetal55} give B0~V but \citet{Garmetal82} 
  	give O9~V. New data obtained by C. Robert et al. (private 
	communication) yield B0.2~V.
\end{itemize}

\section{Spectral classification}

 	The seventh, eighth, and ninth columns in 
Tables~\ref{GOScatalog_namspcl}~and~\ref{GOSsupplement_namspcl} give the
spectral class, luminosity class, and qualifiers that constitute the spectral
classification of each of the O stars in the catalog. The tenth column in 
Table~\ref{GOScatalog_namspcl} gives the secondary component detected in the 
spectrum, if any, while in Table~\ref{GOSsupplement_namspcl} it gives the
spectral type of the WR companion. The eleventh and twelfth columns give the
reference and the table in the reference from which the spectral classification
was obtained.

	The majority of the spectral classifications in the Catalog were 
reported by \citet{Walb72,Walb73,Walb82}. They were performed with photographic
material, specifically with dispersions of $60-80$~\AA\ mm$^{-1}$ and
widening of 1.2~mm on IIa-O plates, obtained at the Kitt Peak, Cerro
Tololo, and Las Campanas Observatories.  These observational parameters
provided an information content about twice that of the classical MK system,
which led to several classification developments discussed by Walborn (1971a).  
For the O stars, these developments included new spectral types at the
extremes of the range, and a second (luminosity-class) dimension
throughout the entire range (the MK system was one-dimensional earlier
than type O9.) 

	As in the MK system, the horizontal (spectral-type) criteria consist of
ratios of \ion{He}{1} and \ion{He}{2} absorption lines, the principal ratio
$\lambda$4471/$\lambda$4541, having a value of unity at
type O7.  A new earliest spectral type of O3 was defined by the
disappearance of the \ion{He}{1} line with these observational characteristics, 
whereas it was weak but well defined in the MK O4 standards (Walborn 1971b).  
At the cool end of the sequence, a new spectral type of O9.7 was
interpolated between the prior O9.5 and B0 types, defined by a value near
unity of the line ratio \ion{Si}{3}~$\lambda$4552/\ion{He}{2}~$\lambda$4541.  
The new luminosity classification was based upon the behavior of \ion{He}{2}
$\lambda$4686 and the \ion{N}{3} $\lambda\lambda 4634-4640-4642$ triplet.
O stars with both of these features in emission had previously been
denoted as Of.  With the higher information-content material, spectra
were recognized with the \ion{He}{2} line neutralized or weakly in absorption 
and \ion{N}{3} emission of intermediate strength, called O(f); and with strong
absorption in the \ion{He}{2} line and weak \ion{N}{3} emission, called O((f)).
These effects were hypothesized to correspond to luminosity differences,
with the full Of stars being the normal supergiants, as was confirmed by 
subsequent calibration work in the papers cited above.  A property of the 
more luminous O3 spectra was selective emission in \ion{N}{4} $\lambda$4058 
stronger than the \ion{N}{3}, which was denoted by the f* qualifier.  A further
development was the recognition of spectra with anomalies in the relative
strengths of CNO lines, categorized as ON or OC \citep{Walb71c,Walb76}.  
These and other notational developments are summarized in 
Table~\ref{qualifiers}. 

	Subsequently, all spectral classification work has shifted to more
powerful digital detectors, which offer greater quantum efficiency, linear 
response, and sky subtraction.  A digital atlas of normal, blue-violet
OB spectra was presented by \citet{WalbFitz90}, and of
peculiar spectra by \citet{WalbFitz00}.  The earliest
spectral types were re-investigated with high-quality digital data by
\citet{Walbetal02b}, resulting in the subdivision of the former O3 type
into three, including the new types O2 and O3.5.  The principal
classification criterion for them is the \ion{N}{4}/\ion{N}{3} selective
emission-line ratio, as explained in detail in that work.   

\section{Astrometry}

	The coordinates of the stars in the catalog are given in 
Tables~\ref{GOScatalog_astrom1}~and~\ref{GOSsupplement_astrom1}. The first two
columns there identify the star, the next two give the epoch J2000.0 right 
ascension and declination, the fifth column gives the coordinate source, and 
the sixth and seventh columns the longitude and latitude. 

	The Tycho-2 catalog \citep{Hogetal00a} was used as our primary
reference for positions, given its precision (median astrometric errors of 7
mas for stars with $V_T < 9$), completeness (99\% of all stars with 
$V \lesssim 11$), and uniformity. 359 of our stars were found in the Tycho-2
catalog. Of those, 336 were present in the main catalog and had proper motion
measurements (these are the stars with the most precise positions) and 14 were 
present in the main catalog but did not have their proper motion measured. The 
remaining 9 Tycho stars were found in the first supplement (7 with proper 
motions and 2 without them) and their coordinates were precessed to epoch 
J2000.0.

	For the 19 stars in the GOS catalog not included in \citet{Hogetal00a},
we obtained their positions from other sources. Eleven of them
were measured using a WFPC2 field which included Tycho-2 stars (used as a
reference to correct the absolute astrometry), three were measured using 
ground-based images, one was listed in the USNO-A2.0 catalog
\citep{Moneetal98}, and four were obtained from the literature (see notes for
Table~\ref{GOScatalog_astrom1}). Those coordinates are expected to have in
general a lower precision than the Tycho ones.

	The eighth and ninth columns in
Tables~\ref{GOScatalog_astrom1}~and~\ref{GOSsupplement_astrom1} give the
parallax ($\pi$) and its uncertainty ($\sigma_{\pi}$) for those stars with 
Hipparcos measurements \citep{ESA97}. The vast majority of the O stars are 
beyond the range in which individual Hipparcos measurements are useful
(i.e. most of the measurements suffer from Lutz-Kelker-type biases, 
\citealt{LutzKelk73}), as evidenced by the large fraction of values in 
Tables~\ref{GOScatalog_astrom1}~and~\ref{GOSsupplement_astrom1} with large
relative uncertainties ($\varepsilon_{\pi}\equiv \sigma_{\pi}/\pi$) or negative
parallaxes. 

	Nevertheless, 24 O stars do have significant Hipparcos parallaxes and
those are shown in Table~\ref{GOScatalog_astrom2}. For those stars, we apply
the method developed by \citet{Maiz01a} to calculate distances using
trigonometric parallaxes. The
method has two parts: First, a vertical (with respect to the Galaxy) spatial
distribution is calculated in a self-consistent way using Hipparcos parallaxes
taking into account Lutz-Kelker-type biases. Second, the probability 
distribution as a function of distance along the line line of sight $r$ is 
calculated using the expression:

\begin{equation}
p(r) = A\, r^2\,e^{-\frac{1}{2}\left(\frac{1-r\pi}
{r\sigma_\pi}\right)^2} \rho(r), \label{dist1}
\end{equation}

\noindent where $\rho(r)$ is the spatial distribution along the line of sight
derived from the vertical spatial distribution and $A$ is a normalizing 
constant. \citet{Maiz01a} used this method to measure the vertical spatial
distribution of Galactic O-B5 stars in the solar neighborhood and found that
it could be well fitted by a combination of a self-gravitating isothermal disk
and a parabolic halo. Here we shall assume that the O stars in this catalog have
the same vertical distribution\footnote{It is not expected that O-B5 stars will 
have a very different vertical spatial distribution compared to O stars except 
maybe for a different fraction of runaway stars \citep{Ston91}. However, since the 
\citet{Maiz01a} method separates the spatial distribution of each type (runaway 
and non-runaway) and here we can identify the category to which each star
belongs, even a possible variation in that fraction is not a problem.}
and that we are not probing distances so 
large that Galactic radial variations in that distribution become important. 
Furthermore, we shall assume that the O-B5 halo is composed of runaway stars 
and the disk of non-runaway ones and assign each star to its corresponding 
parent population with its underlying spatial distribution. It is important to 
stress that:

\begin{itemize}
  \item The most likely value of the distance $r$ to a star (which we shall
  	call $d$) IS NOT $1/\pi$. A 
  	Lutz-Kelker-type correction has to be applied to $1/\pi$ before 
	obtaining $d$ \citep{Smit03}. Also, the correction itself depends on the
	spatial distribution of the parent population (see Eq.~\ref{dist1}), 
	which in our case is chosen from a disk or a halo projected along the 
	line of sight.
  \item The probability distribution for the distance modulus $5\log r-5$ 
	is given by $p(r)$ multiplied by $r$ and renormalized
	to unity, i.e.:

\begin{equation}
p^{\prime}(5\log r-5) = B\, r^3\,e^{-\frac{1}{2}\left(\frac{1-r\pi}
{r\sigma_\pi}\right)^2} \rho(r). \label{dist2}
\end{equation}
	
        We define $DM$ to be the most likely value of $p^{\prime}(5\log r-5)$.
        NEITHER $p(r)$ NOR $p^{\prime}(5\log r-5)$ ARE Gaussian (or even 
	symmetric) probability distributions. 
  \item In Table~\ref{GOScatalog_astrom2} the values given 
	for $d$ and $DM$ are the median of $p(r)$ and $p^{\prime}(r)$,
	respectively, and the values given for $\sigma_{d-}$, $\sigma_{d+}$, 
	$\sigma_{DM-}$, and $\sigma_{DM+}$ are defined by:

\begin{equation}
\begin{array}{lclcc}
\displaystyle{\int_{\sigma_{d-}}^{d} p(r) dr} & = &
\displaystyle{\int_{d}^{\sigma_{d+}} p(r) dr} & = & \\
\displaystyle{\int_{\sigma_{DM-}}^{DM} p^{\prime}(5\log r - 5) d(5\log r -5)} & = & 
\displaystyle{\int_{DM}^{\sigma_{DM+}} p^{\prime}(5\log r - 5) d(5\log r -5)} & = & 0.3413 \\
\end{array}
\label{sigmadef}
\end{equation}

	These choices have the properties that
	$DM-\sigma_{DM-} = 5\log(d-\sigma_{d-})-5$, 
	$DM+\sigma_{DM+} = 5\log(d+\sigma_{d+})-5$, and that,
	if $p(r)$ and $p^{\prime}(5\log r-5)$ were Gaussian distributions,
	$\sigma_d = \sigma_{d-} = \sigma_{d+}$, 
	$\sigma_{DM} = \sigma_{DM-} = \sigma_{DM+}$, with $\sigma_d$ and 
	$\sigma_{DM}$ being the standard deviations of each distribution.
\end{itemize}
  
  	In Fig.~\ref{astromplots} we plot $p(r)$ and $p^{\prime}(5\log r-5)$
for the 24 O stars with significant Hipparcos distances.

\section{Photometry}

\subsection{Optical}

	We have included in the GOS catalog three sets of optical photometry:
Tycho-2 ($B_T$ and $V_T$), Johnson ($V$, $B-V$, and $U-B$), and Str\"omgren
($b-y$, $m_1$, and $c_1$). Tycho-2 photometry is listed in 
Tables~\ref{GOScatalog_phot1}~and~\ref{GOSsupplement_phot1} while Johnson and
Str\"omgren photometry appears in 
Tables~\ref{GOScatalog_phot2}~and~\ref{GOSsupplement_phot2}.

	The Tycho-2 photometry is described in \citet{Hogetal00b,Hogetal00a}.
Two passbands, $B_T$ and $V_T$, similar but not identical to Johnson's $B$ and 
$V$, were observed by the star mapper aboard the
astrometric satellite Hipparcos of the European Space Agency between 1989.85
and 1993.21. Photometric typical uncertainties on $V_T$ are 0.013 magnitudes 
for $V_T < 9$ and 0.10 magnitudes for the full catalog (which includes many
fainter stars); individual
uncertainties are provided for the value of $B_T$ and $V_T$ of each star. 
The main advantages of 
Tycho photometry are (a) its full sky coverage (b) by a single instrument 
(c) not affected by atmospheric extinction and (d) with a common reduction
procedure. The GOS catalog includes $V_T$ magnitudes and uncertainties for 359
stars and $B_T$ magnitudes and uncertainties for 356 stars.

	The Johnson $UBV$ \citep{JohnMorg53} and Str\"omgren $uvby$ 
\citep{CrawMand66} systems are the two most common optical
photometric systems. However, as opposed to Tycho-2, there is no uniform survey 
of the full sky using a single instrument that includes the majority of the 
stars in our catalog for either one of those systems. Therefore, to collect the 
Johnson and Str\"omgren photometric data we used the General Catalog of 
Photometric Data, GCPD \citep{Mermetal97}, which includes a good fraction of our 
sample by compiling data from a large number of sources in the literature.
The GCPD lists values for $V$, $B-V$, and $U-B$ in the case of Johnson
photometry and for $b-y$, $m_1=(v-b)-(b-y)$, and $c_1=(u-v)-(v-b)$ in the case 
of Str\"omgren photometry, but contrary to the Tycho-2 case, no uncertainty
estimates are provided. In order to discard incorrect values and to provide
uncertainty estimates, the following procedure was adopted.

\begin{itemize}
  \item Stars with two or more sources for a given system (Johnson or
        Str\"omgren) but with non-identical values in $V$, $B-V$, $U-B$, 
	$b-v$, $m_1$, or $c_1$ had the mean
  	value calculated and the standard deviation used as the uncertainty.
	In those cases where the uncertainty was below a certain critical value
	(0.01 magnitudes for Johnson data, 0.003 magnitudes for Str\"omgren
	data) it was rounded up to that amount to avoid the existence of
	values with unrealistic precision (e.g. most Johnson photometric data
	are given with only two digits after the decimal point, so rounding  
	errors are expected to play some role).
  \item Stars with two or more sources but all of them being identical were 
  	given the same minimum value of the uncertainty described in the
	previous point.
  \item Stars with a single source were given an uncertainty of 0.02 for
  	Johnson data and 0.01 for Str\"omgren data (those values are typical
	of the uncertainties for cases with two or more sources).
  \item Stars with three or more sources (not all identical) and uncertainty
  	greater than 0.10 in $V$ or in one of the five colors were tested for 
	an incorrect single source by eliminating the most discordant value and 
	checking whether the uncertainty was reduced by a factor of two or more.
	If that was the case, the discordant value was indeed removed and the
	new values for the mean and uncertainty were used.
\end{itemize}

	Two further tests were performed to check for inconsistencies in the
photometry due to possible incorrect measurements. 
First, a plot of $V_T-V$ against $B_T-V_T$ was produced for stars
with Tycho-2 and Johnson data (Fig.~\ref{phottest}, upper left) and we searched
for stars that deviated significantly from the general trend in $B_T-V_T$.
Those stars were flagged with an M for magnitude (discrepancy) in the last 
column of Table~\ref{GOScatalog_phot2}. Some
of those cases may be due to real errors in the data but we suspect that at
least some of them could be due to the different spatial resolution of the 
Hipparcos star mapper compared to ground-based data. Second, a similar test was
produced for each pair of Tycho, Johnson, and Str\"omgren photometries using
the respective ``blue'' minus ``visual'' colors ($B_T-V_T$, $B-V$, and $b-y$,
see Fig.~\ref{phottest}) by comparing the individual color differences against
the general color trend. Stars that deviated from at least one of the trends
were flagged with a C for color (discrepancy) in the last column of 
Table~\ref{GOScatalog_phot2} (stars that passed
the test were indicated there with an N for no discrepancy). For this test we
excluded stars in the supplement (the last column in 
Table~\ref{GOSsupplement_phot2} always shows a W for Wolf-Rayet) since
Str\"omgren $b$ data can be strongly affected by the WR lines around 4650~\AA,
thus producing photometric deviations from the general trend which are known to
be real. Indeed, five of the stars without WR companions flagged with a C are Of 
stars (HD 169582, HD 14947, CW CMa, $\zeta$ Pup, 
and HD 152408), where the same process takes place to a lesser degree.

\subsection{Near infrared}

	For the NIR photometry in the GOS catalog we have used the Two Micron
All Sky Survey (2MASS, \citealt{Skruetal97}). The 2MASS generated $JHK_s$ 
point-source photometry with associated uncertainties for the full sky using 
two automated
1.3 m telescopes, one at Mt. Hopkins, U.S.A., and the other at Cerro Tololo,
Chile. The level-1 requirements of the survey specified magnitude limits for
unconfused point sources outside the Galactic Plane of 15.8, 15.1, and 14.3 in 
$J$, $H$, and $K_s$, respectively, though the actual levels were exceeded in 
many cases.

	We have collected the available information from 2MASS and we have 
listed the NIR photometry in 
Tables~\ref{GOScatalog_phot1}~and~\ref{GOSsupplement_phot1}. We have included 
in those tables $J$, $H$, and $K_s$ magnitudes only if two requirements
were satisfied: (1) the 2MASS photometry for that passband had the best quality 
flag (A), indicating a valid measurement with $\rm{SNR} > 10$ and 
$\sigma < 0.10857$; and (2) no other stars in the GOS catalog exist within a
radius of 3\arcsec\ (this last requirement is included to avoid ``mergers'' in
the data). With those restrictions, the GOS catalog includes 332, 339, and 344
stars with $J$, $H$, and $K_s$ data, respectively. The values listed in
Tables~\ref{GOScatalog_phot1}~and~\ref{GOSsupplement_phot1} have been 
corrected for the zero-point offsets which \citet{Coheetal03} measured in order
to make them consistent with a number of other IR photometric systems. The
corresponding uncertainties in the zero-point offsets have been added in
quadrature to the measured ones.

\section{Other information}

\subsection{Group membership}

	The quality of the data regarding membership to a cluster or an OB
association is extremely varied. For nearby stars, the most detailed study is
that of \citet{deZeetal99}, who use a combination of positions, parallaxes, and
proper motions to identify the membership of OB associations in the solar
neighborhood. For stars located at farther distances, information is usually
poorer and many times based only on position and spectroscopic distances (which
themselves may be suspect, see \citealt{Walb02a}). In some cases the only
information present in the literature of the last 40 years is that of the
original definition of OB Association nomenclature by the IAU
\citep{HaffAlte64}. A more complete recent list for O stars is found in
\citet{Masoetal98} but even there a significant fraction of the stars in the
GOS catalog are missing.

	A related question is that of the identification of runaway stars and
of their parent group (be it a cluster or an association), if any. Again, for 
nearby stars a recent study \citep{Hoogetal00} uses Hipparcos data to provide 
quite detailed information but for the majority of the stars in the catalog 
what is known is usually less. There is even a problem of definition as to what 
constitutes a runaway star \citep{Gies87,Ston91}.

	Given the problems described in the above paragraphs, it is clear that
it is not possible at the present stage to build a high-quality uniform listing
of group membership and runaway character for a sample like the one in this
article (the future looks brighter with GAIA, though). Therefore, we decided to
build this part of the GOS catalog with the approach of providing at least a 
tentative assignment for each star, knowing that future works may supersede 
our compilation. We used as a primary reference for group membership 
\citet{Masoetal98} due to the large size of its sample, and we
complemented it with a series of papers on individual clusters
and associations (see footnotes for 
Tables~\ref{GOScatalog_other}~and~\ref{GOSsupplement_other}, where the group
membership information is listed in the ninth column) and by 
\citet{HaffAlte64}. For those stars not referenced in those articles, we used 
the boundaries for associations established by \citet{HaffAlte64} to determine
their membership tentatively (uncertain determinations are indicated by a
colon). In those cases where a star is known to be part of a cluster or
sub-association inside an association, both are listed. In this paper, we
introduce the term of ``Carina Nebula Association'' to refer to the stars
physically associated with that Nebula. Trumpler 14, Trumpler 16, and 
Collinder 228, as well as some stars in between those sub-associations, are 
included in that association (see \citealt{Walb95} for a discussion of the
stellar content of this region). 

	We also list in the tenth column of
Tables~\ref{GOScatalog_other}~and~\ref{GOSsupplement_other} the runaway 
character for each star based on the following recipe. (1) Stars listed as
runaways in \citet{Hoogetal01} or \citet{Masoetal98} have a ``yes'' on that
column. (2) Other stars in clusters/associations or stars in \citet{Masoetal98}
not listed as runaways have a ``no''. (3) Other stars with ``Field:'' in the 
previous column have a ``no:''.

%Plot with latitude distribution for runaways and non-runaways. Incompleteness
%of runaways? See \citet{Ston91}

\subsection{Proximity/Multiplicity}

	We also list in 
Tables~\ref{GOScatalog_other}~and~\ref{GOSsupplement_other} additional
information regarding proximity and multiplicity. The third column there gives
$d_1$, the distance (in arcseconds)
to the nearest star in the Tycho catalog, and the fourth
one gives $d_2$, the distance to the nearest star in the GOS catalog. Columns
five to eight are a summary of the information in Table 1 of
\citet{Masoetal98}. The fifth column is the binary status (SB1O and SB2O
are well-known single-lined or double-lined spectroscopic binaries, SB1? and
SB2? are their uncertain equivalents, C are stars with apparent constant radial 
velocity, U are stars with unknown status, and an E suffix indicates an
eclipsing or ellipsoidal variable), the sixth column gives the number of
astrometric components, and the seventh and eight columns give the 
distances to the nearest and furthest astrometric components.

\subsection{Online information}

	The catalog is available on the WWW at 
{\tt http://www.stsci.edu/\~{}jmaiz/GOSmain.html}. In addition to all the
information in this paper, the online version also links for each star
to other web-based astronomical data services, such as Aladin, MAST,
and VizieR. We plan to use the web version to include corrections and add
upgrades to the GOS catalog.

\acknowledgments

The authors would like to thank Carmelle Robert and Anne Pellerin for providing 
us with unpublished spectrograms 
of three O-star candidates and Rodolfo Barb\'a for helpful
discussions. Support for this work was provided by grant 82280 from the STScI 
DDRF. H\'ector \'A. Galu\'e and Lisa Wei also acknowledge support from the 
STScI Summer Student Program. This research has made use of Aladin, developed 
by CDS, Strasbourg, France.

\begin{figure}
%\centerline{\includegraphics*[width=\linewidth]{distances1.ps}}
\plotone{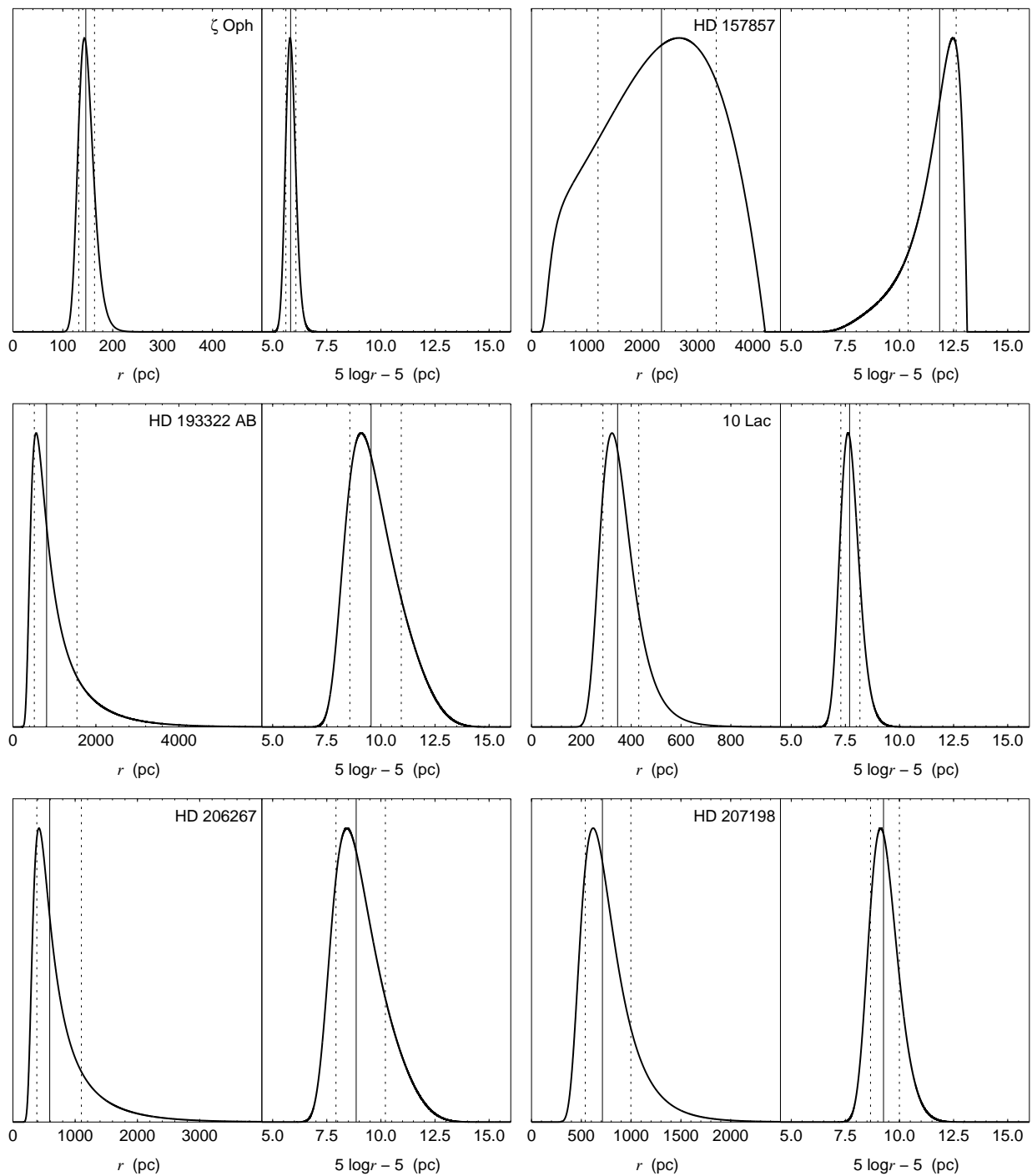}
\caption{$p(r)$ (left plot of each pair) and $p^{\prime}(5\log r-5)$ (right
plot of each pair) for the 24 O stars in the catalog with significant Hipparcos
distances. Solid vertical lines indicate the median value and dashed lines
encompass 68.26\% of the area under each curve (1 $\sigma$ equivalent).}
\label{astromplots}
\end{figure}

\addtocounter{figure}{-1}
\begin{figure}
%\centerline{\includegraphics*[width=\linewidth]{distances2.ps}}
\plotone{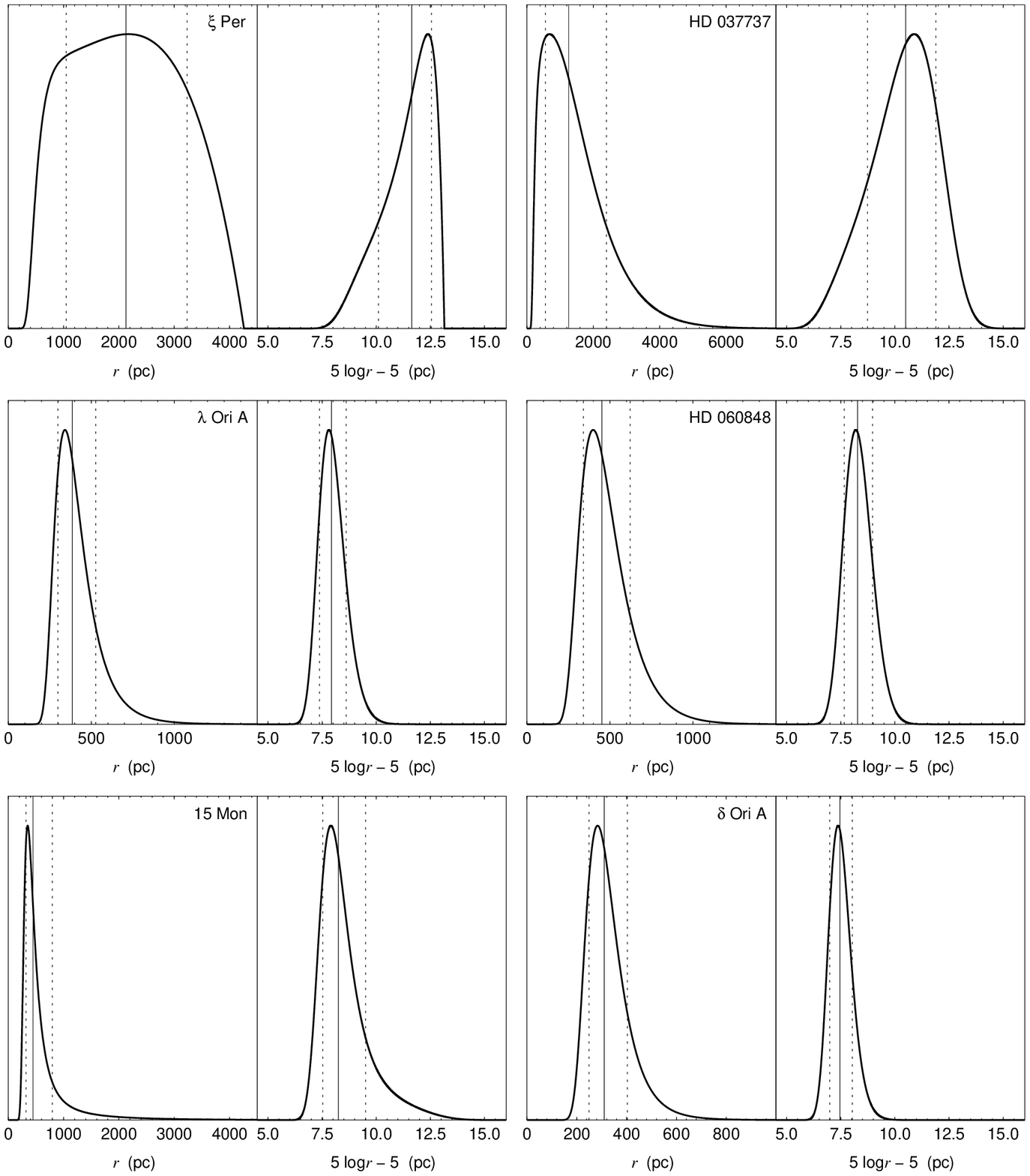}
\caption{(continued)}
\end{figure}

\addtocounter{figure}{-1}
\begin{figure}
%\centerline{\includegraphics*[width=\linewidth]{distances3.ps}}
\plotone{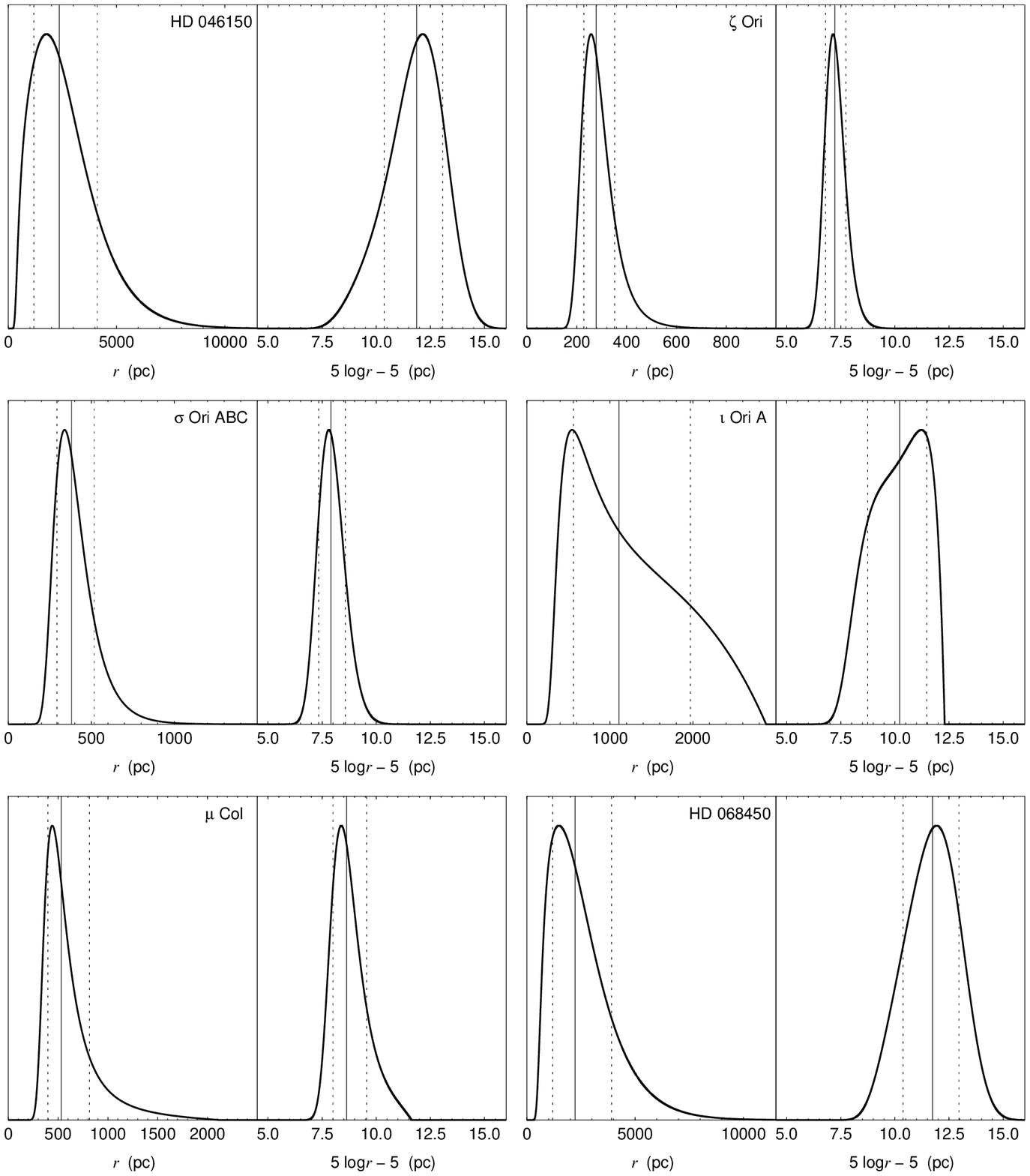}
\caption{(continued)}
\end{figure}

\addtocounter{figure}{-1}
\begin{figure}
%\centerline{\includegraphics*[width=\linewidth]{distances4.ps}}
\plotone{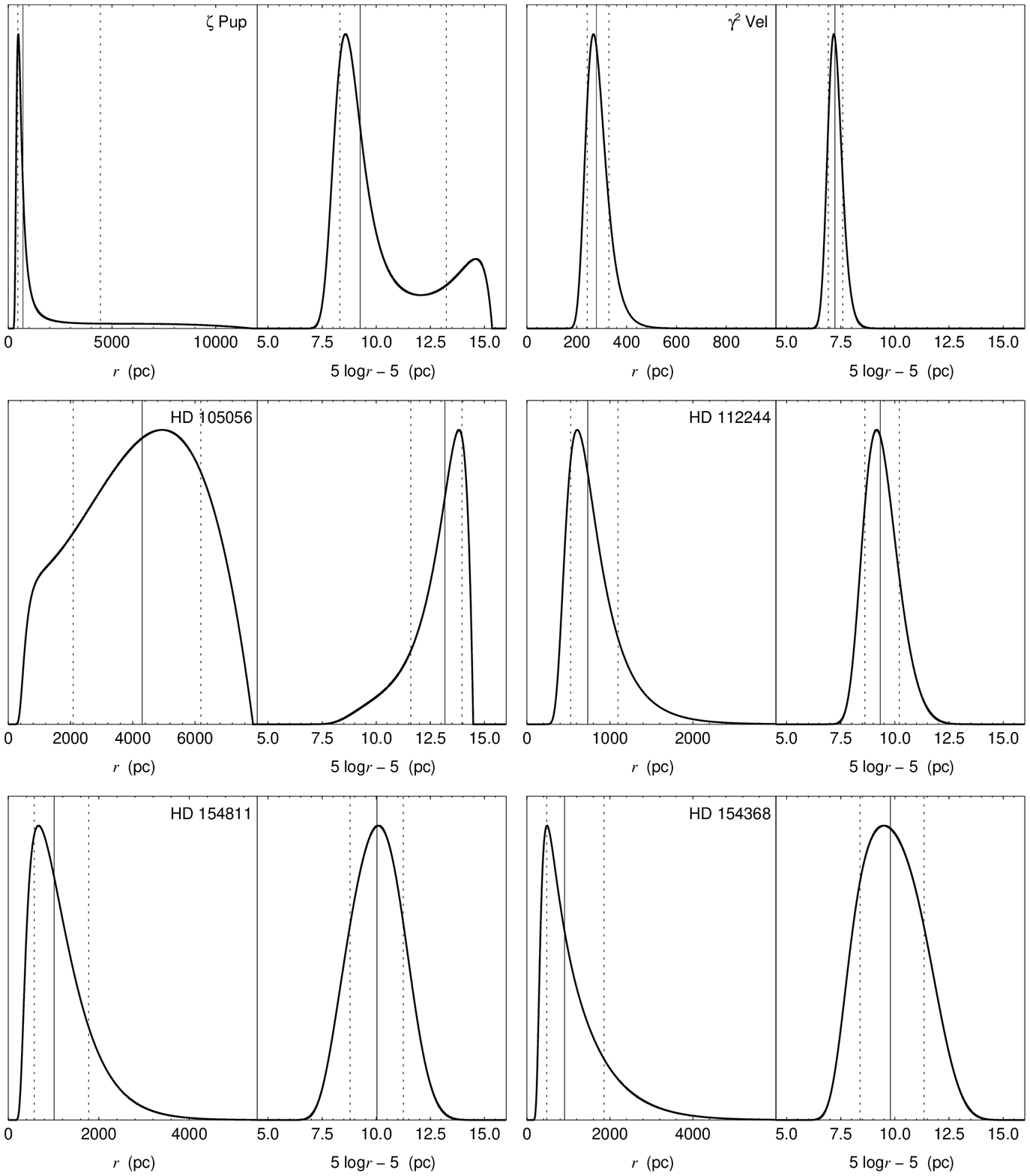}
\caption{(continued)}
\end{figure}

\begin{figure}
%\centerline{\includegraphics*[width=\linewidth]{phottest.ps}}
\plotone{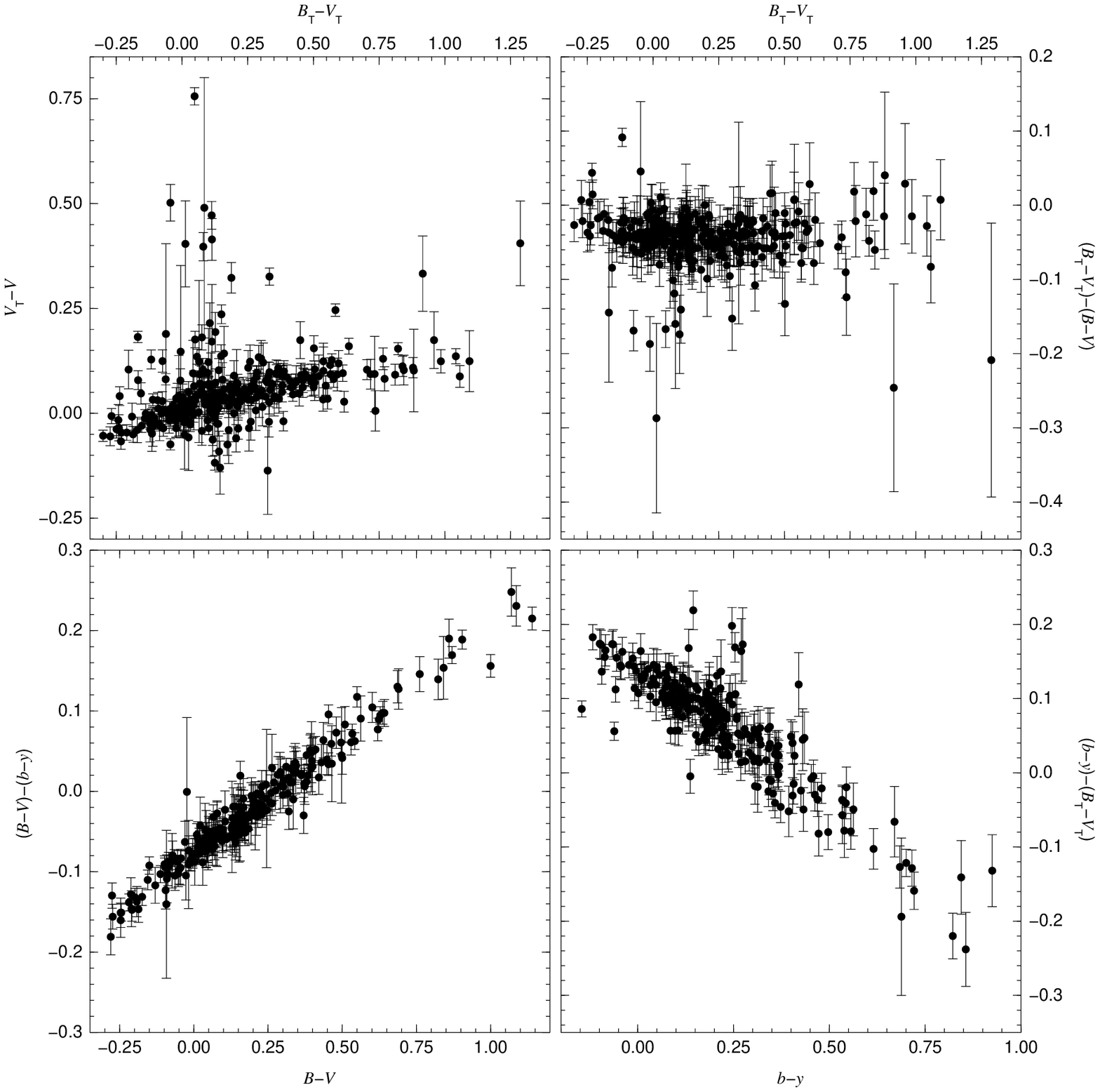}
\caption{Tycho-Johnson (2 upper plots), Johnson-Str\"omgren (lower left plot),
and Str\"omgen-Tycho photometry (lower right plot) checks. 
The upper left plot is used to check
for discrepancies between the measured Tycho and Johnson ``visual'' magnitudes 
while the other three plots are used to check for discrepancies among the
three possible combinations of ``blue'' minus ``visual'' colors using Tycho,
Johnson, and Str\"omgren data.}
\label{phottest}
\end{figure}

\begin{figure}
%\centerline{\includegraphics*[width=\linewidth]{hk_jh.ps}}
\plotone{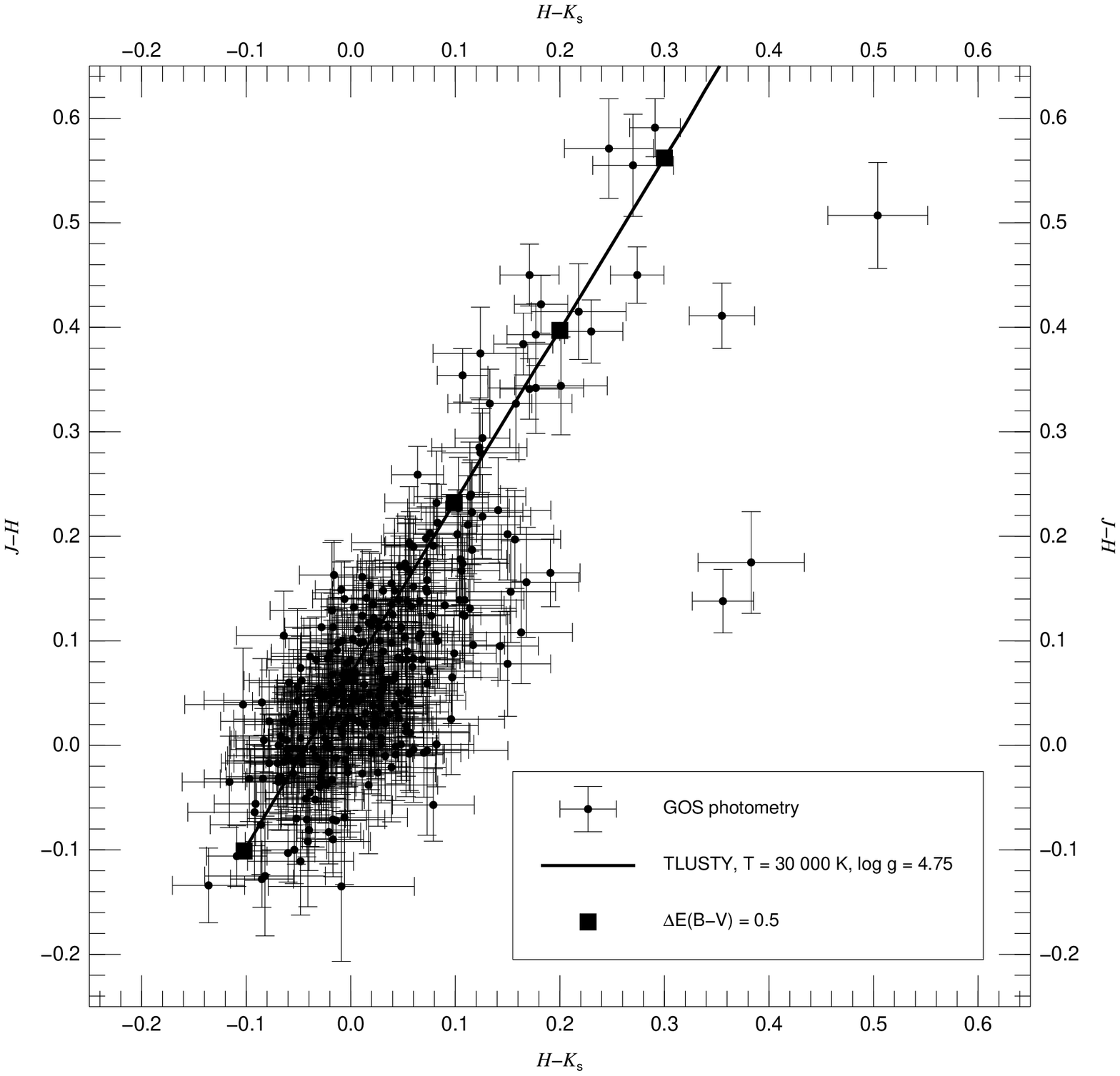}
\caption{$H-K_s$ vs. $J-H$ color-color diagram for the 326 stars in the GOS
catalog with good-quality 2MASS data. The predicted reddened colors using an
$R_V=3.1$ \citet{Cardetal89} extinction law for a $T = 30\, 000$ K, 
$\log g = 4.75$, $Z=Z_\sun$ TLUSTY stellar atmosphere are also shown. The
squares indicate the location of the predicted colors for $E(B-V) = $ 0.0, 
0.5, 1.0, 1.5, and 2.0 (note that the square which corresponds to 
$E(B-V) = 0.5$ is only barely visible).}
\label{2mass}
\end{figure}

\bibliographystyle{apj}
\bibliography{general}

\eject

\addtolength{\oddsidemargin}{-0.6in}
\addtolength{\evensidemargin}{-0.6in}
%\addtolength{\topmargin}{-0.3in}
%\addtolength{\headsep}{+0.6in}
\begin{deluxetable}{rccccclllllc}
\tablecaption{Main catalog: Names and spectral classifications}
\tabletypesize{\scriptsize}
\tablewidth{0pt}
\tablehead{\colhead{GOS ID} & \colhead{HD} & \colhead{comp.} & \colhead{Tycho-2} & \colhead{WDS} & \colhead{other name} & \colhead{SC} & \colhead{LC} & \colhead{qual.} & \colhead{sec.} & \colhead{ref.} & \colhead{tb.} \\
 & & & & & & & & & & & }
\startdata
G$001.91-02.62\_01$ & 164019 & \nodata & 6854-01467-1 & \nodata & \nodata & 
O9.5   & III     & \nodata & \nodata & W82  & 1 \\
G$004.54+00.30\_01$ & 162978 & \nodata & 6845-00692-1 & \nodata & \nodata & 
O7.5   & II      & ((f))     & \nodata & W72  & 2 \\
G$005.82-06.31\_01$ & 168941 & \nodata & 6852-05110-1 & \nodata & \nodata & 
O9.5   & II-III  & \nodata & \nodata & W82  & 1 \\
G$005.97-01.17\_01$ & \nodata & \nodata & \nodata & \nodata & Herschel 36    
 & O7.5   & V       & (n)       & \nodata & W82  & 1 \\
G$006.01-01.20\_01$ & 164794 & \nodata & 6842-01702-1 & \nodata & 
9 Sgr           & O4     & V       & ((f))     & \nodata & W73  & 2 \\
G$006.06-01.20\_01$ & 164816 & \nodata & 6842-01472-1 & \nodata & \nodata & 
O9.5   & III-IV  & (n)       & \nodata & W73  & 2 \\
G$006.12-01.48\_01$ & 165052 & \nodata & 6846-01128-1 & \nodata & \nodata & 
O6.5   & V       & (n)((f))  & \nodata & W73  & 1 \\
G$006.28+23.59\_01$ & 149757 & \nodata & 5632-01042-1 & \nodata & 
$\zeta$ Oph        & O9.5   & V       & nn        & \nodata & L68  & 4 \\
\enddata
\tablerefs{B97: \citet{Burketal97}, D95: \citet{Drisetal95}, G77: \citet{Garretal77}, G82: \citet{Garmetal82}, H99: \citet{HoukSwif99}, L68: \citet{Lesh68}, M76: \citet{MacCBide76}, M98: \citet{Masoetal98}, M55: \citet{Morgetal55}, N88: \citet{NiemMorr88}, P02: \citet{Pennetal02}, W72: \citet{Walb72}, W73: \citet{Walb73}, W76: \citet{Walb76}, W81: \citet{Walb81}, W82: \citet{Walb82}, W00a: \citet{WalbFitz00}, W00b: \citet{WalbHowa00}, W02a: \citet{Walbetal02b}.}
\label{GOScatalog_namspcl}
\end{deluxetable}

\begin{deluxetable}{rccccclllllc}
\tablecaption{Supplement: Names and spectral classifications}
\tabletypesize{\scriptsize}
\tablewidth{0pt}
\tablehead{\colhead{GOS ID} & \colhead{HD} & \colhead{comp.} & \colhead{Tycho-2} & \colhead{WDS} & \colhead{other name} & \colhead{SC} & \colhead{LC} & \colhead{qual.} & \colhead{sec.} & \colhead{ref.} & \colhead{tb.} \\
 & & & & & & & & & & & }
\startdata
G$072.65+02.07\_01$ & 190918 &         & 2682-01564-1 & $20060+3547$ & 
WR 133                 & O9.7   & Iab     & \nodata & WN4      & W74  & 2 \\
G$076.60+01.43\_01$ & 193576 &         & 3151-01250-1 & \nodata & 
V444 Cyg               & O6-8   & III:    & \nodata & WN4      & W74  & 2 \\
G$080.93+04.18\_01$ & 193793 &         & 3164-01678-1 & $20205+4351$ & 
WR 140                 & O4-5   & \nodata & \nodata & WC7      & W90  & 3 \\
G$102.78-00.65\_01$ & 211853 &         & 3986-01670-1 & \nodata & 
GP Cep                 & O8-B0  & III:    & \nodata & WN6      & W74  & 2 \\
G$262.80-07.69\_01$ & 068273 &         & 8140-06533-1 & $08095-4720$ & 
$\gamma^2$ Vel            & O9:    & I:      & \nodata & WC8      & W80  & 5
 \\
G$287.17-00.85\_01$ & 092740 &         & 8626-02024-1 & \nodata & 
WR 22                  & O8-9.5 & III:    & \nodata & WN7      & S99  & 
\nodata \\
G$304.67-02.49\_02$ & 113904 & A       & 8997-02337-1 & $13081-6518$ & 
$\theta$ Mus      A       & O9.7   & Iab     & \nodata & WC6      & W80  & 5
 \\
G$343.49+01.16\_01$ & 152270 &         & 7876-02640-1 & $16543-4149$ & 
WR 79                  & O5-8   & \nodata & \nodata & WC7      & S90  & 1 \\
\enddata
\tablerefs{S90: \citet{Smitetal90}, S99: \citet{Schwetal99}, W74: \citet{Walb74}, W80: \citet{Walb80}, W90: \citet{Willetal90}.}
\label{GOSsupplement_namspcl}
\end{deluxetable}

\begin{deluxetable}{lrlrlrlr}
\tablecaption{Stars with HD numbers in the main catalog and supplement}
\tabletypesize{\scriptsize}
\tablewidth{0pt}
\tablehead{\colhead{HD+c.} & \colhead{GOS ID} & \colhead{HD+c.} & \colhead{GOS ID} & \colhead{HD+c.} & \colhead{GOS ID} & \colhead{HD+c.} & \colhead{GOS ID}}
\startdata
000108 & G$117.93+01.25\_01$ & 041161 & G$164.97+12.89\_01$ & 092206 AB     
 & G$286.22-00.17\_01$ & 099897 & G$293.61-01.28\_01$ \\
001337 & G$117.59-11.09\_01$ & 041997 & G$194.15-01.98\_01$ & 092206 C      
 & G$286.22-00.18\_01$ & 100213 & G$294.81-04.14\_01$ \\
005005 A       & G$123.12-06.24\_01$ & 042088 & G$190.04+00.48\_01$ & 092504
 & G$285.92+00.99\_01$ & 101131 & G$294.78-01.62\_01$ \\
005005 C       & G$123.12-06.24\_02$ & 044811 & G$192.40+03.21\_01$ & 
092740         & G$287.17-00.85\_01$ & 101190 & G$294.78-01.49\_01$ \\
010125 & G$128.29+01.82\_01$ & 045314 & G$196.96+01.52\_01$ & 093027 & 
G$287.61-01.13\_01$ & 101191 & G$294.84-01.68\_01$ \\
012323 & G$132.91-05.87\_01$ & 046056 & G$206.34-02.25\_01$ & 093028 & 
G$287.64-01.19\_01$ & 101205 & G$294.85-01.65\_01$ \\
012993 & G$133.11-03.40\_01$ & 046149 & G$206.22-02.04\_01$ & 093128 & 
G$287.40-00.58\_01$ & 101223 & G$294.81-01.49\_01$ \\
013022 & G$132.91-02.57\_01$ & 046150 & G$206.31-02.07\_01$ & 093129 A      
 & G$287.41-00.57\_01$ & 101298 & G$294.94-01.69\_01$ \\
\enddata
\label{GOShdnumbers}
\end{deluxetable}

\begin{deluxetable}{ll}
\tablecaption{Qualifiers for O spectral types}
\tabletypesize{\scriptsize}
\tablewidth{0pt}
\tablehead{\colhead{Qualifier} & \colhead{Description}}
\startdata
((f))   & Weak\phm{mg}   N\,{\sc iii} $\lambda\lambda 4634-4640-4642$ emission, strong\phm{l or weak.} He\,{\sc ii} $\lambda 4686$ absorption \\
(f)     & Medium\phd\phd N\,{\sc iii} $\lambda\lambda 4634-4640-4642$ emission, neutral or weak        He\,{\sc ii} $\lambda 4686$ absorption \\
f       & Strong\phm{m}  N\,{\sc iii} $\lambda\lambda 4634-4640-4642$ emission, strong\phm{l or weak.} He\,{\sc ii} $\lambda 4686$ emission \\
f?p     & C\,{\sc iii} $\lambda\lambda 4647-51$ emission like N\,{\sc iii} $\lambda\lambda 4634-4640-4642$ \\
\tableline
((f*))  & N\,{\sc iv} $\lambda 4058$ emission $>$ N\,{\sc iii} $\lambda 4640$ emission, strong      He\,{\sc ii} $\lambda 4686$ absorption       (O2-3) \\
(f*)    & N\,{\sc iv} $\lambda 4058$ emission $>$ N\,{\sc iii} $\lambda 4640$ emission, weak\phm{n} He\,{\sc ii} $\lambda 4686$ absorption       (O2-3) \\
f*      & N\,{\sc iv} $\lambda 4058$ emission $>$ N\,{\sc iii} $\lambda 4640$ emission, strong      He\,{\sc ii} $\lambda 4686$ emission\phm{in} (O2-3) \\
\tableline
((f+))  & As       ((f))                   plus Si\,{\sc iv} $\lambda\lambda 4089-4116$ emission (O4-8) \\
\enddata
\label{qualifiers}
\end{deluxetable}

\begin{deluxetable}{llrrlrrrr}
\tablecaption{Main catalog: positions and parallaxes}
\tabletypesize{\scriptsize}
\tablewidth{0pt}
\tablehead{\colhead{GOS ID} & \colhead{other ID} &  \colhead{RA}          & \colhead{dec.}           &  \colhead{ty}                                     &  \colhead{long.}       & \colhead{lat.}           &  \colhead{$\pi$}       & \colhead{$\sigma_{\pi}$} \\  & & \colhead{(J2000)} & \colhead{(J2000)} &  & \colhead{(deg.)} & \colhead{(deg.)} & \colhead{(mas)} & \colhead{(mas)} }
\startdata
G$001.91-02.62\_01$ & HD 164019 & $18\;00\;19.956$ & $-28\;37\;14.66$ & TA & 
$  1.909900$ & $ -2.616621$ & $   1.80$ &   1.57 \\
G$004.54+00.30\_01$ & HD 162978 & $17\;54\;54.042$ & $-24\;53\;13.55$ & TA & 
$  4.536699$ & $  0.301219$ & $  -0.40$ &   0.85 \\
G$005.82-06.31\_01$ & HD 168941 & $18\;23\;25.562$ & $-26\;57\;10.83$ & TA & 
$  5.821025$ & $ -6.312851$ & $  -1.56$ &   1.50 \\
G$005.97-01.17\_01$ & Herschel 36     & $18\;03\;40.200$ & $-24\;22\;43.00$ & 
Va & $  5.970494$ & $ -1.173288$ & \nodata & \nodata \\
G$006.01-01.20\_01$ & 9 Sgr           & $18\;03\;52.446$ & $-24\;21\;38.64$ & 
TA & $  6.008914$ & $ -1.204996$ & $   0.66$ &   1.01 \\
G$006.06-01.20\_01$ & HD 164816 & $18\;03\;56.843$ & $-24\;18\;45.11$ & TA & 
$  6.059112$ & $ -1.195866$ & \nodata & \nodata \\
G$006.12-01.48\_01$ & HD 165052 & $18\;05\;10.551$ & $-24\;23\;54.85$ & TA & 
$  6.121249$ & $ -1.481913$ & $   2.27$ &   1.13 \\
G$006.28+23.59\_01$ & $\zeta$ Oph        & $16\;37\;09.530$ & $-10\;34\;01.75
$ & TB & $  6.281152$ & $ 23.587730$ & $   7.12$ &   0.71 \\
\enddata
\tablecomments{ \\ Al: Secondary component of a multiple system measured with Aladin. \\ Ma: \citet{Massetal01}. \\ NT: Measured from archival NTT data. \\ TA: Main Tycho-2 catalog, columns  5 and  6   (with    proper motion). \\ TB: Main Tycho-2 catalog, columns 28 and 29   (without proper motion). \\ TC: Tycho-2 supplement 1, precessed to J2000.0 with    proper motion. \\ TD: Tycho-2 supplement 1, precessed to J2000.0 without proper motion. \\ US: USNO-A2.0 data. \\ Va: \citet{vanAJone72}. \\ Wa: \citet{Walbetal02b}. \\ WF: WFPC2 data with absolute astrometry corrected using Tycho-2 stars.}
\label{GOScatalog_astrom1}
\end{deluxetable}

\begin{deluxetable}{llrrlrrrr}
\tablecaption{Supplement: positions and parallaxes}
\tabletypesize{\scriptsize}
\tablewidth{0pt}
\tablehead{\colhead{GOS ID} & \colhead{other ID} &  \colhead{RA}          & \colhead{dec.}           &  \colhead{ty}                                     &  \colhead{long.}       & \colhead{lat.}           &  \colhead{$\pi$}       & \colhead{$\sigma_{\pi}$} \\  & & \colhead{(J2000)} & \colhead{(J2000)} &  & \colhead{(deg.)} & \colhead{(deg.)} & \colhead{(mas)} & \colhead{(mas)} }
\startdata
G$072.65+02.07\_01$ & WR 133                 & $20\;05\;57.324$ & $
+35\;47\;18.16$ & TA & $ 72.651611$ & $  2.065109$ & $  -0.96$ &   0.92 \\
G$076.60+01.43\_01$ & V444 Cyg               & $20\;19\;32.423$ & $
+38\;43\;53.97$ & TA & $ 76.597984$ & $  1.428684$ & $   2.92$ &   0.81 \\
G$080.93+04.18\_01$ & WR 140                 & $20\;20\;27.976$ & $
+43\;51\;16.28$ & TA & $ 80.930260$ & $  4.177109$ & $   0.62$ &   0.62 \\
G$102.78-00.65\_01$ & GP Cep                 & $22\;18\;45.604$ & $
+56\;07\;33.90$ & TA & $102.782509$ & $ -0.653186$ & $   1.81$ &   0.95 \\
G$262.80-07.69\_01$ & $\gamma^2$ Vel            & $08\;09\;31.950$ & $
-47\;20\;11.72$ & TC & $262.802551$ & $ -7.685818$ & $   3.88$ &   0.53 \\
G$287.17-00.85\_01$ & WR 22                  & $10\;41\;17.518$ & $
-59\;40\;36.88$ & TA & $287.174194$ & $ -0.847533$ & $   0.61$ &   0.52 \\
G$304.67-02.49\_02$ & $\theta$ Mus      A       & $13\;08\;07.155$ & $
-65\;18\;21.51$ & TA & $304.674561$ & $ -2.490682$ & $   0.03$ &   0.67 \\
G$343.49+01.16\_01$ & WR 79                  & $16\;54\;19.703$ & $
-41\;49\;11.57$ & TA & $343.487335$ & $  1.163930$ & $  -0.59$ &   0.91 \\
\enddata
\tablecomments{ \\ TA: Main Tycho-2 catalog, columns  5 and  6   (with    proper motion). \\ TC: Tycho-2 supplement 1, precessed to J2000.0 with    proper motion.}
\label{GOSsupplement_astrom1}
\end{deluxetable}

\begin{deluxetable}{llcrrrrcc}
\tablecaption{Stars with significant Hipparcos distances}
\tabletypesize{\scriptsize}
\tablewidth{0pt}
\tablehead{\colhead{GOS ID} & \colhead{other ID} &          \colhead{model}       &                                     \colhead{$d$}         & \colhead{$\sigma_{d-}$}  &           \colhead{$\sigma_{d+}$}  &          \colhead{$DM       $} & \colhead{$\sigma_{DM         -}$} &  \colhead{$\sigma_{DM         +}$} \\ 
 & & & \colhead{(pc)} & \colhead{(pc)} & \colhead{(pc)} & & & }
\startdata
G$006.28+23.59\_01$ & $\zeta$ Oph        & runaway &  146 &    14 &    17 & 
 5.83 \phm{xx} &  0.22 &  0.24 \\
G$012.97+13.31\_01$ & HD 157857 & runaway & 2349 &  1147 &   989 & 11.85
 \phm{xx} &  1.46 &  0.76 \\
G$078.10+02.78\_01$ & HD 193322 AB      & disk &  811 &   294 &   731 &  9.55
 \phm{xx} &  0.98 &  1.40 \\
G$096.65-16.98\_01$ & 10 Lac          & disk &  345 &    58 &    84 &  7.70
 \phm{xx} &  0.41 &  0.48 \\
G$099.29+03.74\_01$ & HD 206267 & disk &  590 &   206 &   508 &  8.86
 \phm{xx} &  0.93 &  1.35 \\
G$103.14+06.99\_01$ & HD 207198 & disk &  711 &   172 &   285 &  9.26
 \phm{xx} &  0.60 &  0.73 \\
G$160.37-13.11\_01$ & $\xi$ Per          & runaway & 2125 &  1077 &  1107 & 
11.64 \phm{xx} &  1.53 &  0.91 \\
G$173.46+03.24\_01$ & HD 037737 & disk & 1257 &   698 &  1137 & 10.50
 \phm{xx} &  1.76 &  1.40 \\
\enddata
\label{GOScatalog_astrom2}
\end{deluxetable}

\begin{deluxetable}{llrrrrrrrrrr}
\tablecaption{Main catalog: Tycho and 2MASS photometry}
\tabletypesize{\scriptsize}
\tablewidth{0pt}
\tablehead{\colhead{GOS ID} & \colhead{other ID} & \colhead{$B_T$} & \colhead{$\sigma_{B_T}$} & \colhead{$V_T$} & \colhead{$\sigma_{V_T}$} & \colhead{$J$}   & \colhead{$\sigma_{J}$}   & \colhead{$H$}   & \colhead{$\sigma_{H}$}   & \colhead{$K_s$} & \colhead{$\sigma_{K_s}$} \\ 
 & & & & & & & & & & & }
\startdata
G$001.91-02.62\_01$ & HD 164019 &  9.566 & 0.022 &  9.368 & 0.023 &  8.818 & 
0.025 &  8.762 & 0.026 &  8.813 & 0.025 \\
G$004.54+00.30\_01$ & HD 162978 &  6.194 & 0.014 &  6.188 & 0.010 &  5.970 & 
0.021 &  5.947 & 0.037 &  5.971 & 0.016 \\
G$005.82-06.31\_01$ & HD 168941 &  9.373 & 0.022 &  9.371 & 0.025 &  9.193 & 
0.025 &  9.186 & 0.022 &  9.234 & 0.022 \\
G$005.97-01.17\_01$ & Herschel 36     & \nodata & \nodata & \nodata & \nodata
 &  7.939 & 0.027 &  7.432 & 0.043 &  6.928 & 0.022 \\
G$006.01-01.20\_01$ & 9 Sgr           &  5.942 & 0.014 &  5.960 & 0.010 & 
 5.748 & 0.027 &  5.729 & 0.037 &  5.748 & 0.025 \\
G$006.06-01.20\_01$ & HD 164816 &  7.033 & 0.015 &  7.076 & 0.011 &  7.007 & 
0.022 &  7.034 & 0.045 &  7.089 & 0.019 \\
G$006.12-01.48\_01$ & HD 165052 &  6.938 & 0.015 &  6.874 & 0.011 &  6.484 & 
0.022 &  6.455 & 0.041 &  6.492 & 0.025 \\
G$006.28+23.59\_01$ & $\zeta$ Oph        &  2.575 & 0.014 &  2.546 & 0.009 & 
\nodata & \nodata & \nodata & \nodata & \nodata & \nodata \\
\enddata
\label{GOScatalog_phot1}
\end{deluxetable}

\begin{deluxetable}{llrrrrrrrrrr}
\tablecaption{Supplement: Tycho and 2MASS photometry}
\tabletypesize{\scriptsize}
\tablewidth{0pt}
\tablehead{\colhead{GOS ID} & \colhead{other ID} & \colhead{$B_T$} & \colhead{$\sigma_{B_T}$} & \colhead{$V_T$} & \colhead{$\sigma_{V_T}$} & \colhead{$J$}   & \colhead{$\sigma_{J}$}   & \colhead{$H$}   & \colhead{$\sigma_{H}$}   & \colhead{$K_s$} & \colhead{$\sigma_{K_s}$} \\ 
 & & & & & & & & & & & }
\startdata
G$072.65+02.07\_01$ & WR 133                 &  6.892 & 0.014 &  6.784 & 0.010
 &  6.320 & 0.021 &  6.296 & 0.019 &  6.264 & 0.022 \\
G$076.60+01.43\_01$ & V444 Cyg               &  8.573 & 0.017 &  8.070 & 0.012
 &  6.698 & 0.026 &  6.501 & 0.039 &  6.344 & 0.021 \\
G$080.93+04.18\_01$ & WR 140                 &  7.379 & 0.015 &  6.926 & 0.010
 &  5.548 & 0.019 &  5.410 & 0.024 &  5.054 & 0.017 \\
G$102.78-00.65\_01$ & GP Cep                 &  9.444 & 0.019 &  9.064 & 0.017
 &  7.782 & 0.021 &  7.580 & 0.039 &  7.430 & 0.022 \\
G$262.80-07.69\_01$ & $\gamma^2$ Vel            &  1.627 & 0.007 &  1.793 & 
0.005 & \nodata & \nodata & \nodata & \nodata & \nodata & \nodata \\
G$287.17-00.85\_01$ & WR 22                  &  6.472 & 0.014 &  6.401 & 0.010
 &  5.706 & 0.029 &  5.559 & 0.039 &  5.406 & 0.019 \\
G$304.67-02.49\_02$ & $\theta$ Mus      A       &  5.610 & 0.014 &  5.655 & 
0.009 &  5.572 & 0.024 &  5.547 & 0.048 &  5.451 & 0.019 \\
G$343.49+01.16\_01$ & WR 79                  &  6.833 & 0.015 &  6.625 & 0.011
 &  5.964 & 0.020 &  5.789 & 0.045 &  5.406 & 0.025 \\
\enddata
\label{GOSsupplement_phot1}
\end{deluxetable}

\begin{deluxetable}{llrrrrrrrrrrrrr}
\tablecaption{Main catalog: UBV and Str\"omgen photometry}
\rotate
\tabletypesize{\scriptsize}
\tablewidth{0pt}
\tablehead{\colhead{GOS ID} & \colhead{other ID} & \colhead{$V$} & \colhead{$\sigma_{V}$} & \colhead{$B-V$} & \colhead{$\sigma_{B-V}$} & \colhead{$U-B$} & \colhead{$\sigma_{U-B}$} & \colhead{$b-y$} & \colhead{$\sigma_{b-y}$} & \colhead{$m_1$} & \colhead{$\sigma_{m_1}$}   & \colhead{$c_1$} & \colhead{$\sigma_{c_1}$}   & \colhead{D} \\ 
 & & & & & & & & & & & & & & }
\startdata
G$001.91-02.62\_01$ & HD 164019 & $ 9.278$ & $0.028$ & $ 0.237$ & $0.021$ & 
$-0.713$ & $0.015$ & \nodata & \nodata & \nodata & \nodata & \nodata & \nodata
 & N \\
G$004.54+00.30\_01$ & HD 162978 & $ 6.194$ & $0.016$ & $ 0.035$ & $0.015$ & 
$-0.877$ & $0.017$ & $ 0.096$ & $0.004$ & $ 0.016$ & $0.006$ & $-0.127$ & 
$0.013$ & N \\
G$005.82-06.31\_01$ & HD 168941 & $ 9.356$ & $0.018$ & $ 0.045$ & $0.017$ & 
$-0.835$ & $0.039$ & $ 0.111$ & $0.010$ & $ 0.008$ & $0.010$ & $-0.109$ & 
$0.010$ & N \\
G$005.97-01.17\_01$ & Herschel 36     & \nodata & \nodata & \nodata & \nodata
 & \nodata & \nodata & \nodata & \nodata & \nodata & \nodata & \nodata & 
\nodata & \nodata \\
G$006.01-01.20\_01$ & 9 Sgr           & $ 5.966$ & $0.011$ & $ 0.018$ & 
$0.012$ & $-0.906$ & $0.026$ & $ 0.092$ & $0.008$ & $ 0.003$ & $0.003$ & 
$-0.128$ & $0.008$ & N \\
G$006.06-01.20\_01$ & HD 164816 & $ 7.089$ & $0.013$ & $ 0.003$ & $0.010$ & 
$-0.873$ & $0.022$ & \nodata & \nodata & \nodata & \nodata & \nodata & \nodata
 & N \\
G$006.12-01.48\_01$ & HD 165052 & $ 6.872$ & $0.010$ & $ 0.098$ & $0.012$ & 
$-0.838$ & $0.016$ & $ 0.150$ & $0.010$ & $-0.009$ & $0.010$ & $-0.103$ & 
$0.010$ & N \\
G$006.28+23.59\_01$ & $\zeta$ Oph        & $ 2.566$ & $0.010$ & $ 0.018$ & 
$0.010$ & $-0.860$ & $0.010$ & $ 0.086$ & $0.003$ & $ 0.015$ & $0.005$ & 
$-0.066$ & $0.008$ & C \\
\enddata
\label{GOScatalog_phot2}
\end{deluxetable}

\begin{deluxetable}{llrrrrrrrrrrrrr}
\tablecaption{Supplement: UBV and Str\"omgen photometry}
\rotate
\tabletypesize{\scriptsize}
\tablewidth{0pt}
\tablehead{\colhead{GOS ID} & \colhead{other ID} & \colhead{$V$} & \colhead{$\sigma_{V}$} & \colhead{$B-V$} & \colhead{$\sigma_{B-V}$} & \colhead{$U-B$} & \colhead{$\sigma_{U-B}$} & \colhead{$b-y$} & \colhead{$\sigma_{b-y}$} & \colhead{$m_1$} & \colhead{$\sigma_{m_1}$}   & \colhead{$c_1$} & \colhead{$\sigma_{c_1}$}   & \colhead{D} \\ 
 & & & & & & & & & & & & & & }
\startdata
G$072.65+02.07\_01$ & WR 133                 & $ 6.793$ & $0.057$ & $ 0.134$
 & $0.026$ & $-0.767$ & $0.018$ & $ 0.086$ & $0.008$ & $ 0.160$ & $0.071$ & 
$-0.211$ & $0.028$ & W \\
G$076.60+01.43\_01$ & V444 Cyg               & $ 8.000$ & $0.010$ & $ 0.515$
 & $0.021$ & $-0.385$ & $0.010$ & \nodata & \nodata & \nodata & \nodata & 
\nodata & \nodata & W \\
G$080.93+04.18\_01$ & WR 140                 & $ 6.875$ & $0.021$ & $ 0.395$
 & $0.035$ & $-0.325$ & $0.049$ & \nodata & \nodata & \nodata & \nodata & 
\nodata & \nodata & W \\
G$102.78-00.65\_01$ & GP Cep                 & $ 9.027$ & $0.012$ & $ 0.390$
 & $0.010$ & $-0.550$ & $0.010$ & \nodata & \nodata & \nodata & \nodata & 
\nodata & \nodata & W \\
G$262.80-07.69\_01$ & $\gamma^2$ Vel            & $ 1.798$ & $0.020$ & 
$-0.240$ & $0.020$ & $-0.970$ & $0.020$ & $-0.431$ & $0.010$ & $ 0.671$ & 
$0.010$ & $-0.457$ & $0.010$ & W \\
G$287.17-00.85\_01$ & WR 22                  & $ 6.420$ & $0.022$ & $ 0.078$
 & $0.010$ & $-0.830$ & $0.010$ & $-0.007$ & $0.012$ & $ 0.287$ & $0.018$ & 
$-0.229$ & $0.026$ & W \\
G$304.67-02.49\_02$ & $\theta$ Mus      A       & $ 5.525$ & $0.058$ & 
$-0.027$ & $0.018$ & $-0.854$ & $0.029$ & $-0.117$ & $0.006$ & $ 0.390$ & 
$0.008$ & $-0.321$ & $0.022$ & W \\
G$343.49+01.16\_01$ & WR 79                  & $ 6.615$ & $0.091$ & $ 0.170$
 & $0.054$ & $-0.527$ & $0.033$ & $-0.697$ & $0.066$ & $ 1.700$ & $0.033$ & 
$-0.810$ & $0.033$ & W \\
\enddata
\label{GOSsupplement_phot2}
\end{deluxetable}

\begin{deluxetable}{llrrlrrrlcl}
\tablecaption{Main catalog: Other information}
\tabletypesize{\scriptsize}
\tablewidth{0pt}
\tablehead{\colhead{GOS ID} & \colhead{other ID} & \colhead{$d_1$} & \colhead{$d_2$} & \colhead{SB} & \colhead{mult.} & \colhead{$d_{\rm min}$} & \colhead{$d_{\rm max}$} & \colhead{group} & \colhead{runaway} & \colhead{ref.} \\
 &  & \colhead{(\arcsec)} & \colhead{(\arcsec)} &  &  & \colhead{(\arcsec)} & \colhead{(\arcsec)} &  &  & }
\startdata
G$001.91-02.62\_01$ & HD 164019 &  49.3 & $>100$ & \nodata & \nodata & \nodata
 & \nodata & Field:                   & no: & M04 \\
G$004.54+00.30\_01$ & HD 162978 & $>100$ & $>100$ & C      & 1  & \nodata & 
\nodata & Sgr OB1                  & no & M98 \\
G$005.82-06.31\_01$ & HD 168941 & $>100$ & $>100$ & \nodata & \nodata & 
\nodata & \nodata & Field:                   & no: & M04 \\
G$005.97-01.17\_01$ & Herschel 36     & \nodata & $>100$ & \nodata & \nodata
 & \nodata & \nodata & NGC 6530,Sgr OB1         & no & H64,W82 \\
G$006.01-01.20\_01$ & 9 Sgr           & $>100$ & $>100$ & SB2?   & 1  & 
\nodata & \nodata & Sgr OB1                  & no & M98 \\
G$006.06-01.20\_01$ & HD 164816 & $>100$ & $>100$ & SB2?   & 1  & \nodata & 
\nodata & Sgr OB1                  & no & M98 \\
G$006.12-01.48\_01$ & HD 165052 & $>100$ & $>100$ & SB2O   & 1  & \nodata & 
\nodata & Sgr OB1                  & no & M98 \\
G$006.28+23.59\_01$ & $\zeta$ Oph        & $>100$ & $>100$ & C      & 1  & 
\nodata & \nodata & Sco OB2                  & yes & H01 \\
\enddata
\tablerefs{D99: \citet{deZeetal99}, D95: \citet{Drisetal95}, F58: \citet{Feas58}, G92: \citet{GarmSten92}, H64: \citet{HaffAlte64}, H77: \citet{Heis77}, H93: \citet{Hilletal93}, H01: \citet{Hoogetal01}, L80: \citet{LevaMala80}, M98: \citet{Masoetal98}, M95: \citet{Massetal95}, M01: \citet{Massetal01}, M53: \citet{Morgetal53}, W72: \citet{Walb72}, W73: \citet{Walb73}, W82: \citet{Walb82}, W95: \citet{Walb95}, W02a: \citet{Walbetal02b}, W02b: \citet{Walb02a}, M04: This paper.}
\label{GOScatalog_other}
\end{deluxetable}

\begin{deluxetable}{llrrlrrrlcl}
\tablecaption{Supplement: Other information}
\tabletypesize{\scriptsize}
\tablewidth{0pt}
\tablehead{\colhead{GOS ID} & \colhead{other ID} & \colhead{$d_1$} & \colhead{$d_2$} & \colhead{SB} & \colhead{mult.} & \colhead{$d_{\rm min}$} & \colhead{$d_{\rm max}$} & \colhead{group} & \colhead{runaway} & \colhead{ref.} \\
 &  & \colhead{(\arcsec)} & \colhead{(\arcsec)} &  &  & \colhead{(\arcsec)} & \colhead{(\arcsec)} &  &  & }
\startdata
G$072.65+02.07\_01$ & WR 133                 &  11.3 & $>100$ & SB2O   & 8  & 
6.6   & 38.3  & NGC 6871,Cyg OB3         & no & H64,M98 \\
G$076.60+01.43\_01$ & V444 Cyg               & $>100$ & $>100$ & \nodata & 
\nodata & \nodata & \nodata & Cyg OB1                  & no & M95 \\
G$080.93+04.18\_01$ & WR 140                 & $>100$ & $>100$ & SB2O   & 2 
 & 5.8   & 5.8   & Field                    & no & M98 \\
G$102.78-00.65\_01$ & GP Cep                 & $>100$ & $>100$ & \nodata & 
\nodata & \nodata & \nodata & Cep OB1                  & no & H64 \\
G$262.80-07.69\_01$ & $\gamma^2$ Vel            &  41.2 & $>100$ & SB2O   & 5 
 & 1.6   & 93.5  & Vel OB2                  & no & M98 \\
G$287.17-00.85\_01$ & WR 22                  & $>100$ & $>100$ & \nodata & 
\nodata & \nodata & \nodata & Carina Nebula As.        & no & W95,M04 \\
G$304.67-02.49\_02$ & $\theta$ Mus      A       &   5.3 &  5.5 & \nodata & 
\nodata & \nodata & \nodata & Cen OB1                  & no & M98 \\
G$343.49+01.16\_01$ & WR 79                  &  25.2 & 57.8 & SB2O   & 2  & 
4.3   & 4.3   & NGC 6231,Sco OB1         & no & L80,M98 \\
\enddata
\tablerefs{H64: \citet{HaffAlte64}, L80: \citet{LevaMala80}, M98: \citet{Masoetal98}, M95: \citet{Massetal95}, W95: \citet{Walb95}, M04: This paper.}
\label{GOSsupplement_other}
\end{deluxetable}

\addtolength{\oddsidemargin}{+0.6in}
\addtolength{\evensidemargin}{+0.6in}
%\addtolength{\topmargin}{+0.3in}
%\addtolength{\headsep}{-0.6in}

\end{document}